\newif\ifconfver
\confvertrue        

\ifconfver
	\documentclass{article}
	\usepackage{spconf}
	\title{Symbol-Level Precoding is Symbol-Perturbed ZF When
		Energy Efficiency
		is Sought}
	\name{Yatao Liu and Wing-Kin Ma
		}
	\address{
	Department of Electronic Engineering, The Chinese University of Hong Kong,
 	Hong Kong SAR, China \\
	E-mail: ytliu@ee.cuhk.edu.hk, wkma@ee.cuhk.edu.hk }
\else
	\documentclass[11pt]{article}
	\usepackage{fullpage}
	\title{Symbol-Level Precoding is Symbol-Perturbed ZF When
		Energy Efficiency
		is Sought}
	\author{Wing-Kin Ma}
\fi

\usepackage{calc,amsfonts,amssymb,amsmath,bm,url,color,theorem,graphicx,cite}
\usepackage{psfrag,subfigure,float}
\usepackage{algorithm}
\usepackage{algorithmic}

\definecolor{orange}{RGB}{255,107,0}

\newtheorem{Fact}{Fact}

\newtheorem{Prop}{Proposition}

\theorembodyfont{\rmfamily}

\newcommand\bw{\ensuremath{{\bm w}}}
\newcommand\bW{\ensuremath{{\bm W}}}

\newcommand\bx{\ensuremath{{\bm x}}}

\newcommand\bh{\ensuremath{{\bm h}}}
\newcommand\bH{\ensuremath{{\bm H}}}

\newcommand\bR{\ensuremath{{\bm R}}}

\newcommand\bc{\ensuremath{{\bm c}}}
\newcommand\ba{\ensuremath{{\bm a}}}

\newcommand\bd{\ensuremath{{\bm d}}}
\newcommand\bD{\ensuremath{{\bm D}}}
\newcommand\bu{\ensuremath{{\bm u}}}

\newcommand\beeta{\ensuremath{{\bm \eta}}}

\newcommand\bs{\ensuremath{{\bm s}}}

\newcommand{\Cbb}{\mathbb{C}}

\newcommand{\setR}{\mathcal{R}}

\newcommand{\setS}{\mathcal{S}}

\newcommand{\CN}{\mathcal{CN}}
\newcommand{\Exp}{\mathbb{E}}
\newcommand{\jj}{\mathfrak{j}}
\newcommand{\dec}{\mathrm{dec}}
\newcommand{\Diag}{\mathrm{Diag}}

\newcommand{\eps}{\varepsilon}
\newcommand{\bzero}{{\bm 0}}

\begin{document}

\bibliographystyle{IEEEtran}

\ifconfver
	\ninept
\fi

\maketitle
\begin{abstract}
{
This paper considers symbol-level precoding (SLP) for multiuser multiple-input single-output (MISO) downlink.
SLP is a nonlinear precoding scheme that utilizes symbol constellation structures.
It has been shown that SLP can outperform the popular linear beamforming scheme.
In this work we reveal a hidden connection between SLP and linear beamforming.
We show that under an energy minimization design, SLP is equivalent to a zero-forcing (ZF) beamforming scheme with perturbations on symbols.
This identity gives new insights and they are discussed in the paper.
As a side contribution, this work also develops a symbol error probability (SEP)-constrained SLP design formulation under quadrature amplitude modulation (QAM) constellations.
}
\end{abstract}

%
%
%
%
%
%
%
%
%
%
%
%
%
%
%
%
%
%
%
%


\section{Introduction}

In multiuser MIMO downlink scenarios, linear precoding or beamforming is arguably the most widely used physical-layer transceiver scheme.
Linear beamforming is simple in terms of transceiver structures, and
it has been found in numerous studies that linear beamforming is effective in improving system performance such as total throughput.
Recently there has been interest in another class of precoding schemes called constructive interference, directional modulation, or symbol-level precoding (SLP) \cite{masouros2009dynamic,masouros2011correlation,alodeh2015constructive,spano2016per,alodeh2016energy,kalantari2016directional,alodeh2017symbol,DBLP:journals/corr/KalantariTSCMO17,masouros2015exploiting}.
For convenience, we will use the name SLP when we refer to such schemes.
SLP leverages on the fact that transmitted symbols are drawn from a constellation, such as quadrature amplitude modulation (QAM) and $M$-ary phase shift keying (MPSK), in real world.
By contrast, in linear beamforming, we usually take a level of abstraction from the symbol level.
To be specific, when we design linear beamformers, it is common to adopt quality-of-service (QoS) performance metrics such as the signal-to-interference-and-noise ratio (SINR), achievable rate, and symbol mean square error (MSE).
The use of such metrics frees us from symbol level details and allows us to directly work on the higher level problem of beamforming optimization.
On the other hand, such an
abstraction also precludes
utilization of symbol constellation structures that appear in all practical digital communication systems.

The early idea of SLP dates back to the early 2010 under the name of constructive interference \cite{masouros2009dynamic,masouros2011correlation,alodeh2015constructive}.
There, the rationale is to consider a symbol-dependent linear beamforming scheme in which interference is beneficially aligned at the symbol level.
This requires exploitation of the underlying constellation structures,
and an SLP design for one constellation (e.g., QAM) can be different from that of another (e.g., MPSK).
In recent studies, this rationale is gradually shifting toward a more general philosophy, where SLP is regarded as a generally nonlinear precoder that takes an optimization form.
It is worthwhile to mention that, coincidently, the same precoding philosophy is also seen in the concurrent developments of constant envelope precoding \cite{mohammed2012single,pan2014constant} and one-bit MIMO precoding \cite{jacobsson2017quantized}.

The principles of SLP and linear beamforming are, in essence, different.
In this paper, we draw a connection between the two.
Simply speaking, we show that if we seek to minimize the total transmission energy in the SLP design,
the resulting SLP is equivalent to a zero-forcing (ZF) beamforming scheme with perturbations on symbols.
This result gives new insights into SLP.
We derive the above result based on a symbol error probability (SEP)-constrained SLP design formulation for QAM constellations, which, as a side contribution, is developed in this paper.
After the submission of this work, it has come to our attention that the symbol-perturbed ZF scheme we mentioned above was independently developed in \cite{krivochiza2017low}.
However, the work in \cite{krivochiza2017low} neither noticed nor showed the equivalent relationship of SLP and symbol-perturbed ZF.
In our work we also study a less explored issue in the existing SLP literature, which is about block-level optimization of symbol gains at the user side and will be discussed in Section 5.
Incorporating this issue into the design gives rise to interesting insights as we will illustrate through simulations.

\section{Background}

We consider a multiuser multiple-input single-output (MISO) downlink scenario where a multi-antenna base station (BS) serves $K$ single-antenna users.
The channels from the BS to the users are assumed to be frequency-flat block faded.
Under such settings, the received signals of the users over one transmission block may be modeled as
\begin{equation} \label{eq:sig_mod}
y_{i,t} = \bh_i^H \bx_t + v_{i,t}, \quad
 i=1,\ldots,K, \quad
 t=1,\ldots,T.
\end{equation}
Here, $y_{i,t}$ is the received signal of user $i$ at symbol time $t$;
$\bx_t \in \Cbb^N$ is the multi-antenna transmitted signal from the BS at symbol time $t$, with
$N$ being the number of transmit antennas at the BS;
$\bh_i \in \Cbb^N$ represents the MISO channel from the BS to user~$i$;
$T$ is the transmission block length;
$v_{i,t}$ is noise and we assume $v_{i,t} \sim \CN(0,\sigma_v^2)$ where $\sigma_v^2$ denotes the noise variance.
Assuming perfect channel state information (CSI) at the BS,
the task is to send symbol streams, one designated for one user, via a pertinent MIMO precoding scheme.

Let us briefly recall how the above task is done in conventional linear beamforming.
Let $\{ s_{i,t} \}_{t=1}^T$ be a symbol stream of user $i$.
The transmitted signal $\bx_t$ of linear beamforming takes the form
\begin{equation} \label{eq:linbf}
\bx_t = \sum_{i=1}^K \bw_i s_{i,t},
\end{equation}
where $\bw_i \in \Cbb^N$ is the beamformer associated with the $i$th symbol stream.
There are numerous ways to design the beamformers \cite{Bengtsson2001,schubert2004solution,wiesel2006linear,gershman2010convex,shi2011iteratively},
although the QoS performance metrics used often fall into several types.
In particular, it is common to adopt the SINR
\[
{\sf SINR}_i \triangleq \frac{ \rho_i  | \bh_i^H \bw_i |^2}{ \sum_{j \neq i} \rho_j | \bh_i^H \bw_j |^2 + \sigma_v^2},
\]
where $\rho_i =  \Exp[ | s_{i,t} |^2 ] $, \footnote{Note that in arriving at the SINR expression, we have made two mild assumptions, namely, that i) every stream $\{ s_{i,t} \}_{t=1}^T$ is independent and identically distributed (i.i.d.) with mean zero and variance $\rho_i =  \Exp[ | s_{i,t} |^2 ] $, and ii) one stream is statistically independent of another stream.}
as the QoS performance metric.
The SINR takes a level of abstraction from the symbol level:
it evaluates interference by means of average power, and consequently the underlying constellation structures are not exploited.
A popular beamforming formulation under the SINR metric is the following SINR-constrained design:
\begin{equation} \label{eq:P0}
\begin{aligned}
\min_{\bw_1,\ldots, \bw_K} & ~ \textstyle \Exp[ \| \bx_t \|_2^2 ] = \sum_{i=1}^K \rho_i \|  \bw_i \|^2_2 \\
{\rm s.t.} & ~ {\sf SINR}_i \geq \gamma_i, \quad i=1,\ldots,K,
\end{aligned}
\end{equation}
where $\gamma_i > 0, i=1,\ldots,K,$ are pre-specified SINR requirements;
see the literature \cite{Bengtsson2001,gershman2010convex} for further description.

\section{SEP-Constrained Symbol-Level Precoding}

In this section we consider SLP.
Unlike linear beamforming, which restricts the transmitted signal $\bx_t$ to take the linear form  \eqref{eq:linbf}, SLP
allows $\bx_t$ to be any vector (in  $\Cbb^N$).
It aims at finding an appropriate $\bx_t$ such that desired symbols are shaped at the user side.
To be more specific, we intend to achieve, as accurately as possible,
\begin{equation} \label{eq:approx}
\bh_i^H \bx_t \approx d_i s_{i,t},
\quad \text{for all $i,t$,}
\end{equation}
for some given signal gain factors $d_1,\ldots, d_K > 0$;
$s_{i,t}$'s are again the symbols.
In doing so, we also incorporate other design considerations such as energy efficiency.
Several design formulations for SLP have been proposed in previous works  \cite{alodeh2015constructive,spano2016per,alodeh2016energy,kalantari2016directional,alodeh2017symbol,DBLP:journals/corr/KalantariTSCMO17,masouros2015exploiting},
and in this work we are interested in an SEP-constrained formulation.
In the SEP-constrained formulation, we seek to to minimize the total transmission energy in an instantaneous sense, and, at the same time, we must guarantee the SEP of every user to be no worse than a pre-specified value.
Mathematically,
this is formulated as an optimization problem
\begin{equation} \label{eq:P1}
\begin{aligned}
\min_{ \bx_t } & ~ \textstyle \| \bx_t \|_2^2 \\
{\rm s.t.} & ~ {\sf SEP}_{i,t} \leq \eps_i,  \quad i=1,\ldots,K,
\end{aligned}
\end{equation}
where ${\sf SEP}_{i,t}$ denotes the symbol error probability of $s_{i,t}$ given $s_{i,t}$, which we will define and characterize later, and $\eps_i$'s are pre-specified SEP requirements.
We should mention that SLP requires solving optimization problems on a per-symbol basis, whilst linear beamforming usually solves an optimization problem once for the whole transmission block  (cf. \eqref{eq:linbf}--\eqref{eq:P0}).

Let us characterize the SEP, which depends on the constellation and the detection process at the user side.
We assume that the symbol stream $\{ s_{i,t} \}_{t=1}^T$ of user $i$ is drawn from a QAM constellation
\[
\setS_{i} = \{ s_R + \jj s_I  \mid  s_R, s_I \in \{ \pm 1, \pm 3, \ldots, \pm (2L_i - 1) \} \},
\]
where $\jj = \sqrt{-1}$, and $L_i$ is a positive integer;
note that the constellation size is $4L_i^2$.
Also, we assume that every user has access to its corresponding signal gain factor $d_i$.
In practice, this can be made possible by designing the training phase such that users are able to acquire $d_i$'s from the training signals.
With knowledge of $d_i$'s, the users detect their symbol streams by a simple detection process $\hat{s}_{i,t} = \dec_i( y_{i,t}/ d_i )$, where $\dec_i$ denotes the decision function corresponding to $\setS_i$.
To get some insight with what we will see in the SEP derivations,
observe from the signal model \eqref{eq:sig_mod} that
\[
\frac{y_{i,t}}{d_i} = s_{i,t} + \frac{ b_{i,t} + v_{i,t} }{d_i},
\]
where
\[
b_{i,t} = \bh_i^H \bx_t - d_i s_{i,t}
\]
denotes a residual term of the approximation in \eqref{eq:approx}.
Now, define
\[
{\sf SEP}_{i,t} = {\rm Pr}( \hat{s}_{i,t} \neq s_{i,t} \mid s_{i,t} ).
\]
as the (conditional) SEP in \eqref{eq:P1}.
Also, define
\[
\begin{aligned}
{\sf SEP}_{i,t}^R & = {\rm Pr}( \Re(\hat{s}_{i,t}) \neq \Re({s}_{i,t}) \mid s_{i,t} ), \\
{\sf SEP}_{i,t}^I  & = {\rm Pr}( \Im(\hat{s}_{i,t}) \neq \Im({s}_{i,t}) \mid s_{i,t} ),
\end{aligned}
\]
as the conditional SEPs of the real and imaginary parts of $s_{i,t}$, respectively.
As a standard SEP analysis result, one can show that
\begin{equation} \label{eq:SEP_eqs}
\begin{aligned}
{\sf SEP}^R_{i,t} &  =
Q \left(  \frac{\sqrt{2}}{\sigma_v} ( d_i - \Re( b_{i,t}  ) ) \right) +
Q \left(  \frac{\sqrt{2}}{\sigma_v} ( d_i + \Re(  b_{i,t} ) ) \right) \\
& \leq 2 Q \left( \frac{\sqrt{2}}{\sigma_v} ( d_i - |\Re(  b_{i,t} )| ) \right), \qquad | \Re(s_{i,t}) | < 2L_i - 1,  \\
{\sf SEP}^R_{i,t} &  = Q \left( \frac{\sqrt{2}}{\sigma_v} ( d_i + \Re(  b_{i,t} ) ) \right), \qquad \Re(s_{i,t}) =  2L_i - 1,    \\
{\sf SEP}^R_{i,t} &  = Q \left( \frac{\sqrt{2}}{\sigma_v} ( d_i - \Re(  b_{i,t} ) ) \right), \qquad \Re(s_{i,t}) =  -2L_i + 1.
\end{aligned}
\end{equation}
Also, the same result applies to ${\sf SEP}_{i,t}^I$ if we replace
``$R$'' with ``$I$'' and
 ``$\Re$'' with ``$\Im$''.

Our next task is to turn the SEP constraints in \eqref{eq:P1} to a form suitable for optimization.
Let $$\bar{\eps}_i = 1 - \sqrt{1- \eps_i},$$
and observe that
\begin{equation} \label{eq:imply_1}
{\sf SEP}^R_{i,t} \leq \bar{\eps}_i, ~ {\sf SEP}^I_{i,t} \leq \bar{\eps}_i \quad \Longrightarrow \quad {\sf SEP}_{i,t} \leq \eps_i.
\end{equation}
Also, it is shown from \eqref{eq:SEP_eqs} that
\begin{equation} \label{eq:imply_2}
{\sf SEP}^R_{i,t} \leq \bar{\eps}_i   \quad \Longleftarrow \quad
 -d_i + a^R_{i,t} \leq \Re(b_{i,t}) \leq d_i - c^R_{i,t},
\end{equation}
where
\[
\begin{aligned}
a_{i,t}^R & = \left\{
\begin{array}{ll}
\alpha_i, & | \Re(s_{i,t}) | < 2L_i - 1  \\
\beta_i, & \Re(s_{i,t}) = 2L_i - 1 \\
-\infty, & \Re(s_{i,t}) = -2L_i + 1
\end{array}
\right.  \\
c_{i,t}^R & = \left\{
\begin{array}{ll}
\alpha_i, & | \Re(s_{i,t}) | < 2L_i - 1  \\
-\infty, & \Re(s_{i,t}) = 2L_i - 1 \\
\beta_i, & \Re(s_{i,t}) = -2L_i + 1
\end{array}
\right.
\end{aligned}
\]
with
\[
\alpha_i = \frac{\sigma_v}{\sqrt{2}} Q^{-1} \left( \frac{\bar{\eps}_i}{2} \right),
\quad
\beta_i = \frac{\sigma_v}{\sqrt{2}} Q^{-1} \left( \bar{\eps}_i \right),
\]
and that the same result applies to ${\sf SEP}_{i,t}^I$ if we replace
``$R$'' with ``$I$'' and
 ``$\Re$'' with ``$\Im$''.
Using \eqref{eq:imply_1}--\eqref{eq:imply_2}, we obtain the implication
\[
\begin{aligned}
-d_i + a^R_{i,t} & \leq \Re(b_{i,t}) \leq d_i - c^R_{i,t}, \\
-d_i + a^I_{i,t} & \leq \Im(b_{i,t}) \leq d_i - c^I_{i,t}
\end{aligned}
\quad \Longrightarrow \quad
{\sf SEP}_{i,t} \leq \eps_i.
\]
By plugging the above implication into the constraints of Problem~\eqref{eq:P1}, we obtain a tractable SLP design problem.
Let us summarize the result.

\begin{Fact}
The SEP-constrained SLP design problem \eqref{eq:P1} can be handled, in a restrictive sense, by the following problem
\begin{equation} \label{eq:P1_res}
\begin{aligned}
\min_{ \bx_t } & ~ \textstyle \| \bx_t \|_2^2 \\
{\rm s.t.} & ~  -\bd + \ba^R_t \leq \Re( \bH \bx_t - \bD \bs_t )  \leq \bd - \bc^R_t,   \\
   & ~ -\bd + \ba^I_t \leq \Im( \bH \bx_t - \bD \bs_t )  \leq \bd - \bc^I_t,
\end{aligned}
\end{equation}
where $\bH = [~ \bh_1,\ldots, \bh_K ~]^H$,
$\bd = [~ d_1,\ldots, d_K ~]^T$,
$\bD = \Diag(d_1,\ldots,d_K)$,
$\ba^R_t = [~ a_{1,t}^R, \ldots, a_{K,t}^R ~]^T$,
$\ba^I_t = [~ a_{1,t}^I, \ldots, a_{K,t}^I ~]^T$,
$\bc^R_t = [~ c_{1,t}^R, \ldots, c_{K,t}^R ~]^T$,
$\bc^I_t = [~ c_{1,t}^I, \ldots, c_{K,t}^I ~]^T$.
In particular, any feasible solution to Problem~\eqref{eq:P1_res} is a feasible solution to Problem~\eqref{eq:P1}.
Also, Problem~\eqref{eq:P1_res} is a convex quadratic program.
\end{Fact}

In this paper we will focus on Problem~\eqref{eq:P1_res}.
Note that this new formulation is restrictive owing to the inequality in \eqref{eq:SEP_eqs} and the implication in \eqref{eq:imply_1}.
In practice, such restriction is considered mild especially if the SEP requirement $\eps_i$ is small.
Problem~\eqref{eq:P1_res} is similar to the formulation in \cite{DBLP:journals/corr/KalantariTSCMO17} in which a more intuitive idea of ``relaxed decision region'' was introduced to
guarantee certain SNR performance.
Our formulation, in comparison, provides a more precise control on SEP performance guarantees.

\section{Main Result}

Our main result is described as follows.

\begin{Prop} \label{prop:slp_zf}
	Suppose that $\bH$ has full row rank.
	The optimal solution $\bx_t^\star$ to Problem \eqref{eq:P1_res} is given by
	\[
	\bx_t^\star = \bH^\dag ( \bD \bs_t + \bu_t^\star ),
	\]
	where $\bH^\dag = \bH^H (\bH \bH^H )^{-1}$ is the pseudo-inverse of $\bH$,
	and $\bu_t^\star$ is the solution to
	\begin{equation} \label{eq:P1_u}
	\begin{aligned}
	\min_{ \bu_t } & ~ \textstyle ( \bD \bs_t + \bu_t )^H \bR ( \bD \bs_t + \bu_t ) \\
	{\rm s.t.}
	& ~ -\bd + \ba^R_t \leq \Re( \bu_t )  \leq \bd - \bc^R_t, \\
	& ~ -\bd + \ba^I_t \leq \Im( \bu_t )  \leq \bd - \bc^I_t.
	\end{aligned}
	\end{equation}
	with $\bR = (\bH \bH^H )^{-1}$.
	
\end{Prop}

{\em Proof:} \
Note that $\bR$ is nonsingular.
Thus, any $\bx_t \in \Cbb^N$ can be represented by
\[
\bx_t = \bH^H \bR ( \bD \bs_t  + \bu_t ) +  \beeta_t,
\]
for some $\bu_t \in \Cbb^K$, $\beeta_t \in \setR(\bH^H)^\perp$.
Here $\setR(\bH^H)^\perp$ denotes the orthogonal complement of the range space of $\bH^H$.
By the above change of variable,
and using $\bH \beeta_t = \bzero$,
Problem~\eqref{eq:P1_res} can be equivalently transformed to
\begin{equation*}
\begin{aligned}
\min_{ \bu_t \in \Cbb^K, \beeta_t \in \setR(\bH^H)^\perp } & ~ \textstyle \textstyle ( \bD \bs_t + \bu_t )^H \bR ( \bD \bs_t + \bu_t ) + \| \beeta_t \|_2^2 \\
{\rm s.t.}
& ~ -\bd + \ba^R_t \leq \Re( \bu_t )  \leq \bd - \bc^R_t, \\
& ~ -\bd + \ba^I_t \leq \Im( \bu_t )  \leq \bd - \bc^I_t.
\end{aligned}
\end{equation*}
It is seen from the problem above that $\beeta_t = \bzero$ must be true at the optimum.
The proof is complete.
\hfill $\blacksquare$

\medskip

Proposition~\ref{prop:slp_zf} reveals a hidden connection between SLP and linear beamforming:
under the total energy minimization formulation considered above, SLP is equivalent to
a {\em symbol-perturbed ZF beamforming scheme}---which takes the form $\bx_t = \bH^\dag ( \bD \bs_t + \bu_t )$---with the symbol perturbation $\bu_t$ being optimized according to $\bs_t$.
It is also interesting to note the following identity:

\begin{Fact} \label{fac:lin}
	Consider the linear beamforming scheme $\bx_t = \sum_{i=1}^K \bw_i s_{i,t}$.
	Suppose that $\bH$ has full row rank, and that $\bw_i \in \setR(\bH^H)$ for all $i$.
	Then the linear beamforming scheme is equivalent to a symbol-perturbed ZF scheme $\bx_t = \bH^\dag( \bD \bs_t + \bu_t)$ where $\bu_t = (\bH \bW - \bD ) \bs_t$, and $\bW = [~ \bw_1,\ldots,\bw_K ~]$.
\end{Fact}
Fact~\ref{fac:lin} can be easily shown by putting $\bu_t = (\bH \bW - \bD ) \bs_t$ into the symbol-perturbed ZF scheme.
Fact~\ref{fac:lin} suggests that a linear beamforming scheme, under a mild assumption, can be regarded as an instance of symbol-perturbed ZF.
Some further discussions are as follows.
\begin{enumerate}
	\item While the original SLP problem \eqref{eq:P1_res} and its equivalent formulation \eqref{eq:P1_u} are both convex, the latter is easier to handle.
	Problem~\eqref{eq:P1_u} is a quadratic program with bound constraints,
	which has been extensively studied and has efficient solvers readily available \cite{bochkanov2013alglib}.
	
	\item We see from Problem~\eqref{eq:P1_u} that the signal gain factors $d_i$'s also control the bounds of the perturbations $\bu_t$'s. In particular, if we choose $d_i= \alpha_i$ for all $i$, then, for instances where $|s_{i,t}| < 2L_i - 1$ for all $i$, we have $\bu_t = \bzero$ and the SLP scheme reduces to the ZF.

\end{enumerate}

\section{Further Issues}

Thus far, in our development, we have assumed that the signal gain factors $d_i$'s are given.
A question arising is how we may choose $d_i$'s.
An optimal way of doing so is to consider the following problem
\begin{equation} \label{eq:P1_blk}
\begin{aligned}
\min_{ \bx_1, \ldots, \bx_T, \bd } & ~ \textstyle 
\frac{1}{T} \sum_{t=1}^T \| \bx_t \|_2^2 \\
{\rm s.t.} & ~ {\sf SEP}_{i,t} \leq \eps_i,  \quad i=1,\ldots,K, ~  t=1,\ldots,T, \\
& ~ \bd \geq \bzero;
\end{aligned}
\end{equation}
where
we seek to optimize SLP and the signal gain factors simultaneously
by minimizing the total power over the transmission block.
Using Proposition~\ref{prop:slp_zf}, we can recast the problem (in a restrictive sense) as
\begin{equation} \label{eq:P1_u_blk}
\begin{aligned}
\min_{ \bu_1, \ldots, \bu_T, \bd } & ~
\textstyle  \frac{1}{T} \sum_{t=1}^T ( \bD \bs_t + \bu_t )^H \bR ( \bD \bs_t + \bu_t ) \\
{\rm s.t.}
& ~ -\bd + \ba^R_t \leq \Re( \bu_t )  \leq \bd - \bc^R_t, \quad t=1,\ldots,T, \\
& ~ -\bd + \ba^I_t \leq \Im( \bu_t )  \leq \bd - \bc^I_t, \quad t=1,\ldots,T, \\
& ~ \bd \geq \bzero,
\end{aligned}
\end{equation}
The above problem is convex.
We can also consider an alternative formulation wherein the peak energy, rather than the average, is minimized:
\begin{equation}
\begin{aligned} \label{eq:P1_u_blk_alt}
\min_{ \bu_1, \ldots, \bu_T, \bd } & ~
 \max_{t=1,\ldots,T} ( \bD \bs_t + \bu_t )^H \bR ( \bD \bs_t + \bu_t ) \\
{\rm s.t.}
& ~ -\bd + \ba^R_t \leq \Re( \bu_t )  \leq \bd - \bc^R_t, \quad t=1,\ldots,T, \\
& ~ -\bd + \ba^I_t \leq \Im( \bu_t )  \leq \bd - \bc^I_t, \quad t=1,\ldots,T, \\
& ~ \bd \geq \bzero,
\end{aligned}
\end{equation}
The above formulation may be useful when one desires to reduce the energy spread of the transmitted signals over symbol time.
Again, the above problem is convex.

Although the two design problems in \eqref{eq:P1_u_blk} and \eqref{eq:P1_u_blk_alt} are convex, they are by no means easy to deal with.
The reason is
that the number of variables of Problems \eqref{eq:P1_u_blk} and \eqref{eq:P1_u_blk_alt} scales with the block length $T$.
As such, they are large-scale problems when $T$ is large (which is typical in standards), and development of fast algorithms is required.
We leave the latter as an open problem for future work.
In this paper we will use general purpose convex optimization software (such as \texttt{CVX}) to solve Problems~\eqref{eq:P1_u_blk} and \eqref{eq:P1_u_blk_alt},
and our emphasis will be placed on demonstrating performance gains of  Problems~\eqref{eq:P1_u_blk} and \eqref{eq:P1_u_blk_alt} by simulations.

On the other hand, one can use heuristics to determine $\bd$.
Suppose that $L_i > 1$ for all $i$.
Also, let us make a mild assumption that there exist $s_{i,t}$ such that $|s_{i,t}| < 2L_i -1$ for all $i$.
From Problem~\eqref{eq:P1_u}, one can verify that it must hold that
\[
d_i \geq \alpha_i, \quad \text{for all $i$.}
\]
Since using smaller $d_i$'s should be helpful in reducing the total transmission energy, we can choose
\begin{equation} \label{eq:heur}
d_i = \zeta \cdot \alpha_i, \quad i=1,\ldots,K,
\end{equation}
where $\zeta \geq 1$ is a manually chosen scaling factor.

\section{Simulation Results}
In this section, we present simulation results to show the performance of SLP and compare it with other schemes.
In the simulations, we use the total transmission power $\frac{1}{T}\sum_{t=1}^{T}\|\bx_t\|_2^2$ and peak total energy $\max_{t=1,\dots,T}\|\bx_t\|_2^2$ as the performance metrics.
The simulation settings are as follows.
The number of transmit antennas $N = 16$;
the number of users $K = 16$;
the elements $h_{ij}$ of $\bH$ are randomly generated at each trial and follow $\CN(0,1)$ in an i.i.d. manner;
the power of noise is $\sigma_v^2 = 1$;
the transmission block length is $T = 50$;
the symbols $s_{i,t}$'s are uniformly generated from the 16-QAM constellation;
we set $\eps_1 = \cdots = \eps_K = \eps$.
For each simulation scenario, we generate 100 channel realizations to get an average result of the performance metrics.

Here we consider two benchmark schemes: the ZF scheme \cite{Wiesel2008} and optimal linear beamforming scheme \cite{Bengtsson2001}.
The two schemes are designed such that the SEP requirements in \eqref{eq:P1} are satisfied.
For the ZF scheme, it can be verified that $\bx_t^{ZF} = \bH^H(\bH \bH^H )^{-1}\bD \bs_t$ with $d_i= \alpha_i$ for all $i$ achieves the requirements.
For the optimal linear beamforming scheme, we have the following fact:

\begin{Fact} \label{fac:linsep}
    Consider the linear beamforming scheme in \eqref{eq:linbf}-\eqref{eq:P0}.
	Suppose that the multiuser interferences are approximated as complex circular Gaussian random variables.
    The optimal beamforming design in \eqref{eq:P0} guarantees the SEP requirements in \eqref{eq:P1} if we choose
\[
\gamma_i = \frac{\rho_i}{2}\left[Q^{-1}\left(\frac{1-\sqrt{1-\eps_i}}{2}\right)\right]^2, \quad i=1,\ldots,K.
\]
\end{Fact}
We skip the proof of the above fact.
In fact, the result is almost a folklore.

\begin{figure}[ht!]
    \centering
    \includegraphics[width=1\linewidth]{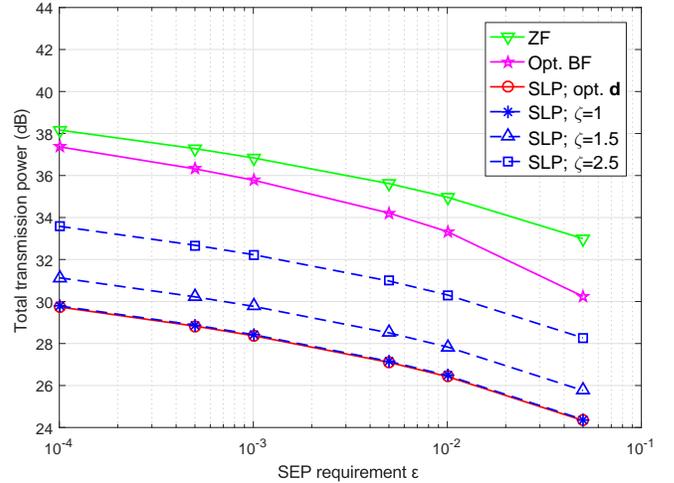}
    \caption{Total transmission power $\frac{1}{T}\sum_{t=1}^{T}\|\bx_t\|_2^2$ with respect to the SEP requirement $\eps$.}
    \label{fig1}
\end{figure}

We first examine total transmission power performance.
We consider the optimal SLP design in Problem~\eqref{eq:P1_u_blk}.
We also try SLP designs using the heuristic choice of $d_i$'s in \eqref{eq:heur}.
The results are shown in Fig.~\ref{fig1}.
As can be seen,
the SLP schemes consume much less power than the two benchmark schemes.
Moreover, we see that the heuristic SLP schemes can also achieve surprisingly good performance.
In particular, when $\zeta = 1$, the performance is almost optimal.
As a future work, it would be interesting to further study why this is so.

\begin{figure}[ht!]
    \centering
    \includegraphics[width=1\linewidth]{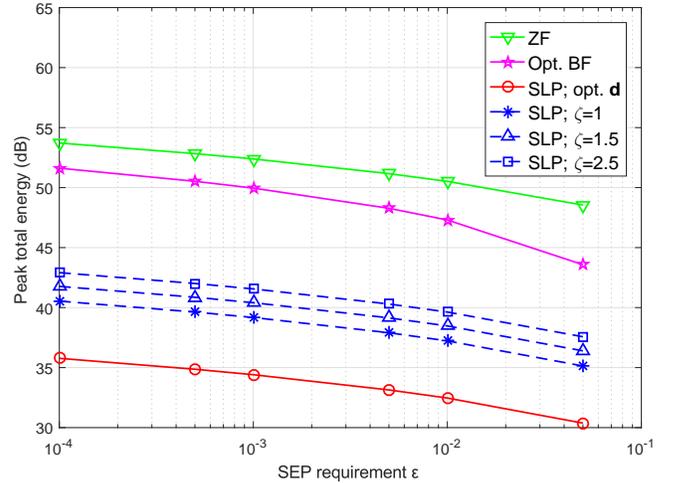}
    \caption{Peak total energy $\max_{t=1,\dots,T}\|\bx_t\|_2^2$ with respect to the SEP requirement $\eps$.}
    \label{fig2}
\end{figure}

Next, we turn to peak total energy performance.
We replace the average power minimization design \eqref{eq:P1_u_blk} with the peak total energy minimization design \eqref{eq:P1_u_blk_alt}.
The previously tested heuristic SLP schemes are also tried.
The results, shown in Fig.~\ref{fig2},
illustrate that the SLP schemes, even the heuristic SLP ones, outperform the benchmark schemes significantly.
However, unlike the previous total power result, there is
a large performance
gap between the optimal SLP and heuristic SLP schemes.
We thus conclude that optimal SLP is powerful in the case of peak total energy minimization.


\section{Conclusions}
In previous works, SLP has been considered as a nonlinear precoding technique.
In this work we showed that SLP can be regarded as a linear ZF scheme with perturbations on symbols.
We also examined new SLP designs and demonstrated their potential by simulations.
Numerical results illustrated that SLP can lead to significant improvement in energy efficiency.

\clearpage
\bibliographystyle{IEEEtran}
\bibliography{refs}

\end{document}